# Ultrashort infrared 2.5 –11µm pulses: spatiotemporal profiles and absolute nonlinear response of air constituents

S. Zahedpour, S. W. Hancock, and H. M. Milchberg

Institute for Research in Electronics and Applied Physics, University of Maryland, College Park, MD 20742

**We measure the detailed spatiotemporal profiles of femtosecond laser pulses in the infrared wavelength range $\lambda = 2.5 - 11\mu m$, and the absolute nonlinear response of major air constituents ($N_2$, $O_2$, and Ar) over this range. The spatiotemporal measurements reveal wavelength-dependent pulse front tilt and temporal stretching in the infrared pulses.**

There is increasing development of intense short pulse laser sources from the mid-infrared (MIR) through the long wave-infrared (LWIR) region. These include nonlinear optics-based sources such as optical parametric amplification (OPA) [1], difference frequency generation (DFG) [2], and optical parametric chirped pulse amplification (OPCPA) [3], as well as solid state gain media [4], and high pressure $CO_2$ gas lasers [5]. The existence of air transparency windows within the MIR-LWIR range [6] has motivated the study of their propagation in atmosphere and applications such as remote sensing in the molecular fingerprint region [7], high harmonic generation [8] and shaped multi-octave supercontinuum generation [9]. Important to all these applications is nonlinear propagation, which depends on the near-instantaneous (electronic) and delayed (rotational, vibrational) nonlinear responses of the medium. For the 50-300fs MIR-LWIR pulses of this experiment, the electronic and rotational responses dominate; the non-resonant Raman vibrational response of $O_2$ and $N_2$ is negligible owing to insufficient optical bandwidth.

In this Article we present measurements of the detailed spatiotemporal profiles of femtosecond laser pulses in the infrared wavelength range $\lambda = 2.5 - 11\mu m$, and the absolute nonlinear response of major air constituents ($N_2$, $O_2$, and Ar) over this range. We also investigate possible resonant 2-photon vibrational excitation of $N_2$ near λ=8µm, which is of interest for high power LWIR laser pulse propagation [10]. The spatiotemporal measurements reveal the wavelength-dependent pulse front tilt and temporal stretching induced by the difference frequency generation scheme used to generate the infrared pulses.

There is a paucity of absolute nonlinear response measurements in the MIR and LWIR. Recent work includes measurements by our group of the nonlinear response of air constituents in the range $\lambda = 0.4 - 2.4\mu m$ [11] using single shot supercontinuum spectral interferometry (SSSI) [12-15], which enables separation of electronic and rotational contributions to the total nonlinearity. For $\lambda \sim 10$ µm pulses, the total electronic plus rotational response of major air constituents and noble gases was measured [16, 17] using four-wave mixing of two rotational lines from a 200ps $CO_2$ laser pulse. In that case, the long pulsewidth and prevented separation of the electronic and rotational contributions.

Here, we use SSSI to measure the space and time-resolved nonlinear phase shift imparted on a weak supercontinuum (SC) probe pulse by the MIR-LWIR pump pulse-induced refractive index shift in the gases studied. As discussed in [11], the extracted nondegenerate nonlinearity coefficients are within ~5% of their degenerate MIR counterparts., with even better agreement in the LWIR. As depicted in Fig. 1, the MIR-LWIR pump at $\lambda_{DF}$ is generated by non-collinear difference frequency generation (DFG, Light Conversion) between ultrashort signal $\lambda_S = 1.1 - 1.6$ µm and idler $\lambda_I = 1.6 - 2.6$ µm pulses in a nonlinear crystal ($AgGaS_2$). The signal and idler were generated by an optical parametric amplifier (OPA, HE-TOPAS Prime, Light Conversion) pumped by 36 fs, 8 mJ pulses centered at λ~806nm from a 1 kHz Ti:Sapphire amplifier system. Supercontinuum (SC) pulses (400-750nm) were generated from filamentation of ~200μJ, 806nm laser pulses from the same laser, focused at ~f/150 in a 2.5 atm xenon gas cell, followed by a Michelson interferometer, which splits the pulse into a SC reference-probe pair separated by ~2.5ps. The SC pair co-propagates with the MIR-LWIR pump pulse into the gas target cell, with the reference pulse in advance of the pump and the probe overlapped with it, and encoded with the pump-induced transient nonlinear phase shift. Polarization of the SC pulses is adjusted by rotating the polarization of the Ti:Sapphire laser pulse entering the Xe cell with a λ/2 plate. Two gas pressure ranges were used in the target gas cell. For pump pulses $\lambda_{DF} = 3.0 - 6.5 \ \mu m$, the cell was filled to 1 atm with the test gases (N$_2$, O$_2$, and Ar). For pump pulses $\lambda_{DF} = 7.0 - 11.0 \ \mu m$, the cell was filled to 42 atm. This was done to increase the signal to noise ratio, as the output of the DFG drops to <10μJ at λ=11μm. The pump focusing lens and entrance window of the gas cell are BaF$_2$ to avoid absorption losses. The focal plane of the reference/probe in the cell interaction region is imaged onto the slit of an imaging spectrometer. The reference and probe interfere in the spectral domain, producing a 2D spectral interferogram $\Delta\Phi(x,\omega)$ (space resolution $x$ along slit, spectral resolution $\omega$ perpendicular to slit) at the spectrometer's focal plane CCD camera. A chopper wheel blocks and unblocks the pump pulse on consecutive shots, enabling subtraction of pump-off background interferograms from pump-on shots. Pump-probe group velocity walk-off is negligible (< 2 fs) in the gas targets of this experiment. In a procedure described previously [12], Fourier analysis of $\Delta\Phi(x,\omega)$, using the measured spectral phase of the probe, gives the 1D-space- and time-resolved phase shift $\Delta\varphi(x,t)$ imposed on the probe by the pump-induced refractive index shift in the test gas. Time and space resolution in these measurements is ~5 fs and ~3μm.

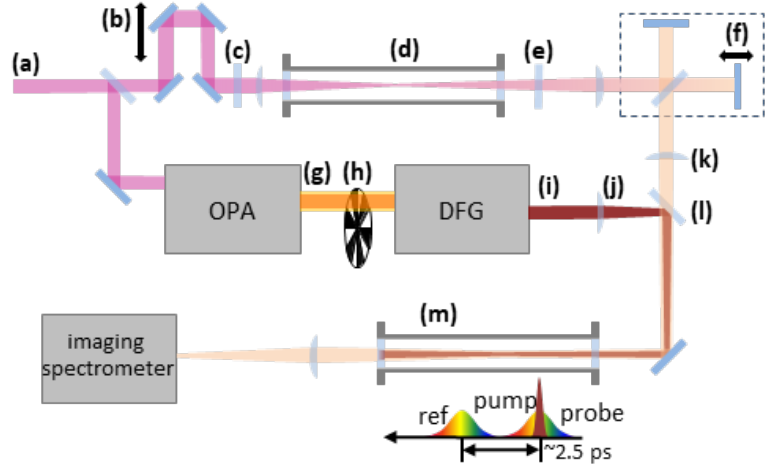

**Fig. 1.** (a) Ti:Sapphire pump, ~806nm, 36fs, 10mJ; 8mJ is split and used to pump OPA; (b) delay line for timing of supercontinuum (SC) reference/probe with respect to MIR-LWIR pump pulse; (c) $\lambda/2$ plate for rotating SC polarization; (d) 2.5 atm xenon SC cell; (e) dichroic mirror, rejects 806nm pump, transmits 400nm-750nm SC pulses; (f) Twin reference and probe SC pulses separated by ~2.5ps produced by Michelson interferometer; (g) signal (1.1-1.6 μm) and idler (1.6-2.6 μm) pulses out of OPA; (h) chopper to block pump for measuring background phase; (i) MIR-LWIR pump out of DFG, (j) BaF$_2$ pump focusing lens (k) BK7 reference/probe SC focusing lens (l) gold dichroic mirror, transmits 400-700nm, reflects 3-12μm (m) test gas cell (up to 42 atm) with BaF$_2$ entrance window, and fused silica exit window to absorb the pump. For each data set, the weak window response induced on the probe is measured by evacuating the cell. Each phase image composed of 1-2×10$^4$ shots.

The refractive index shift $\Delta\varphi(x,t)$ experienced by the probe is the sum of the electronic and rotational responses; if the phase shift imposed on a probe polarized parallel to the pump is $\Delta\varphi_\parallel(x,t) = \Delta\varphi_{elec}(x,t) + \Delta\varphi_{rot}(x,t)$, then the phase shift imposed on a probe polarized perpendicular to the pump is $\Delta\varphi_\perp(x,t) = \Delta\varphi_{elec}(x,t)/3 -$

$\Delta\varphi_{rot}(x,t)/2$ [18]. Solving yields the pure electronic and rotational phase shifts in terms of the ∥ and ⊥ phase shifts: $\Delta\varphi_{elec}(x,t) = 3(\Delta\varphi_\|(x,t) + 2\Delta\varphi_\perp(x,t))/5$ and $\Delta\varphi_{rot}(x,t) = 2(\Delta\varphi_\|(x,t) - 3\Delta\varphi_\perp(x,t))/5$.

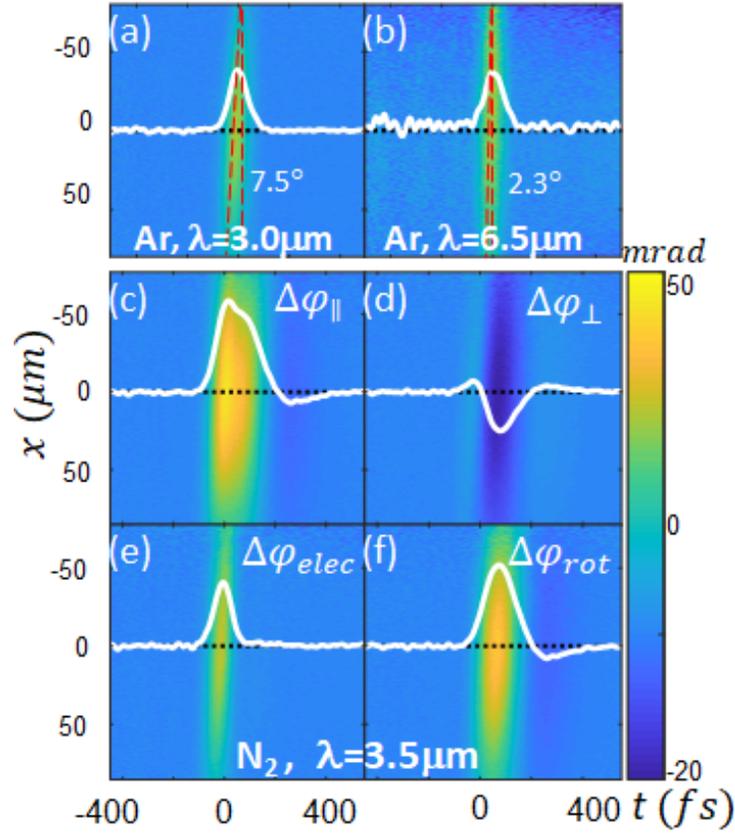

**Fig. 2.** 2D spatiotemporal intensity traces from SSSI. (a) and (b): $\Delta\varphi_{elec}(x,t) \propto I(x,t)$ in Ar for pump pulses at λ=3.0μm and λ=6.5μm, showing pulse front tilt of 7.5° and 2.3°. White curves are central lineouts, with FWHM pulsewidths 77 fs and 92 fs. (c) and (d): probe phase shifts $\Delta\varphi_\|(x,t)$ and $\Delta\varphi_\perp(x,t)$ in $N_2$ perpendicular and parallel to λ=3.5μm pump. (e) and (f): $\Delta\varphi_{elec}(x,t)$ (lineout FWHM = 81 fs) and $\Delta\varphi_{rot}(x,t)$ extracted from $\Delta\varphi_\|$ and $\Delta\varphi_\perp$.

In Figs. 2(a) and 2(b), we first show spatiotemporal traces of MIR-LWIR pulses using Ar as the test gas and ∥ probe polarization. Here, the extracted phase shift imposed on the probe pulse is $\Delta\varphi_\|(x,t) = 2n_2 I(x,t) k_p L$ [15] from the near-instantaneous electronic nonlinearity of Ar (for ⊥ probe polarization, $\Delta\varphi_\perp(x,t)$ $\Delta\varphi_\|(x,t)/3$ for a purely electronic nonlinearity), where $k_p = 2\pi/\lambda_p$ is the probe central wavenumber, $n_2$ is the nonlinear refractive index (electronic), and $L$ is the interaction length, which cancels out in the analysis. The phase shift profiles are therefore a record of the spatiotemporal intensity profile $I(x,t)$. To our knowledge, these are the first direct single shot measurements of spatiotemporal intensity profiles in the MIR-LWIR. It is immediately seen that SSSI reveals pulse front tilt in the DFG-generated pulses, with the effective tilt angle decreasing with increasing λ. The traces also show pulse temporal broadening for increasing λ. The tilt is a consequence of phase matching in non-collinear DFG, and the broadening is dominated by strongly increasing group velocity dispersion (GVD) in our optical materials at long $\lambda_{DF}$.

For $N_2$ test gas, $\Delta\varphi_\|$ and $\Delta\varphi_\perp$ imposed on the probe by a λ=3.5 μm pump in $N_2$ are shown in Figs. 2(c) and (d), and the extracted $\Delta\varphi_{elec}(x,t)$ and $\Delta\varphi_{rot}(x,t)$ are shown in Figs. 2(e) and (f). The negative $\Delta\varphi_\perp$ in Fig. 2(d) occurs from ⊥ probe sampling of molecules whose ensemble average axis alignment is along the pump polarization, which gives a deficit in phase shift compared to the case of random alignment. Notable in Fig. 2 are the similar FWHM pulsewidths of $\Delta\varphi_{elec}$ at λ=3.0μm (77 fs) and 3.5μm (81 fs) in Ar and N$_2$ (for the same pump wavelength they are equal to within measurement error, as expected). At λ= 6.5μm the FWHM is 92 fs.

Extraction of the $n_2$ coefficients from the measured phase shifts $\Delta\varphi_\parallel$ and $\Delta\varphi_\perp$ proceeds as in our prior work [11], where we reference these measurements to the rotational responses in nitrogen and oxygen without explicit need for either gas density, interaction length, or pump intensity profile measurements. The electronic and rotational phase shifts can be written

$$\Delta\varphi_{elec}(x,t) = 2n_2 I_0 k_p L (N/N_0) f(x,t)$$

$$\Delta\varphi_{rot}(x,t) = 2\pi N n_0^{-1} I_0 k_p L \Delta\alpha(\lambda_p)\Delta\alpha(\lambda_{pu}) g(t) * f(x,t)$$

(1)

where $n_2$ and $n_0$ are the nonlinear (electronic) and linear refractive indices at 1 atm, $I_0$ is the peak spatiotemporal intensity, $f(x,t)$ is the normalized intensity envelope, with peak value of 1, $N$ is the molecular density, $N_0$ is the molecular density at 1 atm, $L$ is the interaction length in the gas target, $\Delta\alpha(\lambda_p)$ and $\Delta\alpha(\lambda_{pu})$ are the very weakly frequency-dependent molecular polarizability anisotropy [11, 14] at the probe and pump wavelengths, and the convolution $I_0\Delta\alpha(\lambda_{pu}) g(t) * f(x,t) = I_0\Delta\alpha(\lambda_{pu}) \int_{-\infty}^{t} g(t-t')f(x,t')dt' = \left(<\cos^2\theta>_{x,t} - \frac{1}{3}\right)$ is the ensemble average transient alignment induced by the pump pulse. In the latter expression, we use the rescaled impulse response function for quantized rotations of a rigid molecular rotor [14,15], $g(\tau) = \left(\frac{-16\pi}{15\hbar c}\right) \sum_{j=1}^{\infty} \frac{j(j-1)}{2j-1} \left(\frac{\rho_j^{(0)}}{2j+1} - \frac{\rho_{j-2}^{(0)}}{2j-3}\right) e^{-\gamma_{j,j-2}\tau} \sin\omega_{j,j-2}\tau$ [19], where $\rho_j^{(0)}$ is the thermal population of rotational state $j$, $\omega_{j,j-2} = 4\pi c B(2j-1)$ and $\gamma_{j,j-2}$ are the transition frequency and dephasing rate between states $j$ and $j-2$, and $B$ is the rotational constant of the molecule.

The effect of collisional dephasing on $(t)$, to be discussed later in the paper, is an important concern given that the test gas cell is at 42 atm for $\lambda > 6.5 \mu m$. Expressed directly as a refractive index shift experienced by the probe pulse, $\Delta n_p(x,t) = 2n_2 I(x,t) + \int_{-\infty}^{t} R(t-t')I(x,t')dt'$, where the impulse response function is $R(\tau) = 2\pi N n_0^{-1} \Delta\alpha(\lambda_p)\Delta\alpha(\lambda_{pu}) g(\tau)$. It is clearly seen from the expressions in Eq. (1) how $n_2$ is extracted given the measured 2D datasets $\Delta\varphi_{elec}(x,t)$ and $\Delta\varphi_{rot}(x,t)$, the measured 2D spatiotemporal envelope $f(x,t)$ (from the electronic response), and the known impulse response $g(t)$, with $N$, $I_0$, and $L$ cancelling out. For nitrogen, The dispersion in $\Delta\alpha$ is even weaker in the MIR-LWIR than in our prior case at $\lambda < 2.4 \mu m$ [20,21], with $\Delta\alpha_{N_2}(\lambda_{pu}) \approx \Delta\alpha_{N_2}(0) = 6.6 \times 10^{-25} cm^3$ [11]. For oxygen, we assume $\Delta\alpha(\lambda_p) = \Delta\alpha(\lambda_{pu})$ for all analysis and we use the values of $\Delta\alpha_{O_2} = 10.2 \times 10^{-25} cm^3$ measured in [15].

For each molecular dataset, the analysis proceeds in practice by performing the 2D least squares fit $g(t) * \Delta\varphi_{elec}(x,t) = \mu_1 \Delta\varphi_{rot}(x,t)$ to yield $\mu_1$, where $\Delta\varphi_{elec}$ and $\Delta\varphi_{rot}$ are the 2D SSSI traces. Each SSSI trace has ~100 points in $x$ and ~50–100 points in , enabling $< 10^4$ points for fitting per shot. This gives the best fit value of $n_2 = \mu_1 \pi n_0^{-1} N_0 \Delta\alpha(\lambda_p)\Delta\alpha(\lambda_{pu}) = \mu_1 \pi n_0^{-1} N_0 (\Delta\alpha)^2$. Here, as discussed, $\Delta\alpha$ is taken as spectrally flat, and The Kerr coefficient $n_2$ of argon is measured relative to that of nitrogen using the same pump pulse parameters. Performing a 2D least squares fit for $\mu_2$, $\Delta\varphi_{elec}^{Ar}(x,t) = \mu_2 \Delta\varphi_{elec}^{N_2}(x,t)$, gives $n_2^{Ar} = n_2^{N_2} \mu_2 N_{Ar}/N_{N_2}$ from Eq. (1).

At the longer wavelengths produced by the DFG, the conversion efficiency drops due to the Manley-Rowe relations. In order to maintain the signal-to-noise ratio at an acceptable level for pump wavelengths $\lambda > 6.5 \mu m$, we increased the test gas pressure to 42 atm. At higher pressures, collisional dephasing causes the rotational response function to decay exponentially with time, as seen in the expression for $g(t)$. If we assume a dephasing rate $\gamma_{j,j-2} = \gamma$ independent of transition, then we can exploit the exponential decay of the peaks of the rotational revivals to extract $\gamma$ by fitting $(\Delta\varphi_{rot})_{peaks} \propto e^{-\gamma t}$, to the full and half revivals measured near $t = mT$ and $t = \left(m+\frac{1}{2}\right)T$ for $m = 0 - 4$, where $T = (2cB)^{-1} = 11.6 ps$ is the revival period for $O_2$ [14]. Figure 3 shows the initial $\Delta\varphi_{rot}$ response near $t = 0T$ at $O_2$ gas pressures 1–41 atm, with all peaks normalized to 1, followed by the revivals near $t = \frac{1}{2}T$ and $t = T$. The dashed line shows a decaying exponential fit to the revival curve for 8 atm.

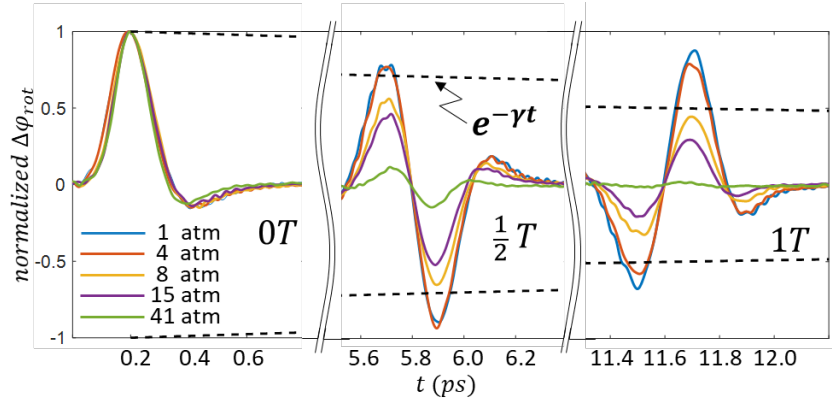

**Fig. 3.** Use of the decay of rotational revival peaks in $\Delta\varphi_{rot}$ for $O_2$ to extract the dephasing rate $\gamma$ as function of gas pressure. The result is $\gamma = 7.4 \times 10^9 (s \cdot atm)^{-1} P(atm)$. The rotational revival period for $O_2$ is $T = 11.6$ ps.

Extracting $\gamma$ as a function of pressure yields the damping rate $\gamma = 7.4 \times 10^9 (s \cdot atm)^{-1} P(atm)$. At $P$=42 atm, $\gamma^{-1} = 3.2$ ps , which is much longer than the maximum pump pulse duration of ~300 fs (see Fig. 4), ensuring negligible effect on the rotational response during the temporal window where $\Delta\varphi_{elec}$ and $\Delta\varphi_{rot}$ are measured.

The results of our measurements and analysis are shown in Table 1. As in our prior results at $\lambda \le 2.4\mu m$ in the mid-IR [11], there is little dispersion in $n_2$. In the region near $\lambda$=8.0μm for $N_2$, which was scanned continuously through 7.5-8.5 μm, we observed no signature of 2-photon resonant absorption [10].

Spatiotemporal traces for pump pulses $\lambda$=7–11μm are shown in Figure 4. As in Fig. 2, all pulses generated by non-collinear DFG show pulse front tilt. This is a natural consequence of the phase matching condition $\mathbf{k}_S = \mathbf{k}_I + \mathbf{k}_{DF}$ and $k_S/n_S = k_I/n_I + k_{DF}/n_{DF}$, where $\mathbf{k}_S$, $\mathbf{k}_I$, and $\mathbf{k}_{DF}$ are the signal, idler, and difference wave wavenumbers; $k_S$, $k_I$, and $k_{DF}$ are their magnitudes; and $n_S$, $n_I$, and $n_{DF}$ are the refractive indices of $AgGaS_2$ at those wavelengths, whose Sellmeier curves are found in [22]. In non-collinear geometry, the output pulse front tilt

**Table 1. Measured values of $n_2$ ($\times 10^{20}$ cm$^2$/ W) (electronic) for major air constituents at atmospheric pressure**

| $\lambda(\mu m)^{(a)}$ | 3.0 | 3.5 | 4.0 | 4.5 | 5.0 | 5.5 | 6.0 | 6.5 | $n_{2,rot}$ |
|---|---|---|---|---|---|---|---|---|---|
| $N_2$ | 7.6±0.5 | 7.3±0.5 | 7.6±0.5 | 7.4±0.5 | 7.2±0.5 | 8.0±0.5 | 7.1±0.5 | 7.8±0.5 | 23.0 |
| $O_2$ | 7.9±0.4 | 8.6±0.5 | 7.7±0.4 | 9.6±0.5 | 7.3±0.4 | 8.1±0.5 | 9.1±0.4 | 8.9±0.5 | 53.3 |
| Ar | 8.9±0.6 | 9.8±0.6 | 8.3±0.5 | 8.5±0.5 | 7.3±0.5 | 8.0±0.5 | 8.9±0.6 | 8.8±0.6 | 0 |
| $\lambda(\mu m)^{(a)}$ | 7.0 | 7.5 | 8.0 | 8.5 | 9.0 | 9.5 | 10.0 | 10.5$^{(c)}$ | $n_{2,eff}(\tau_{opt})^{(b)}$ |
| $N_2$ | 7.8±0.5 | 7.1±0.5 | 7.6±0.5 | 7.8±0.5 | 7.9±0.6 | 8.0±0.5 | 7.6±0.5 | 6.9±0.5 | 32.5 (400fs) |
| $O_2$ | 8.0±0.5 | 8.5±0.5 | 7.2±0.5 | 8.3±0.5 | 7.9±0.5 | 8.8±0.6 | 7.7±0.5 | 7.3±0.5 | 64.5 (480fs) |
| Ar | 7.8±0.5 | 9.2±0.6 | 8.6±0.6 | 9.2±0.6 | 9.2±0.6 | 9.0±0.6 | 7.4±0.6 | 7.7±0.5 | 8.5 (average) |

(a) Pump wavelengths $\lambda_{DF}$ from the DFG are determined by measuring the wavelengths of the OPA signal ($\lambda_S$) and idler ($\lambda_I$) output using a mid-infrared spectrometer (AvaSpec-NIR512-2.5-HSC) spanning 1.0 μm-2.5μm.
(b) The effective nonlinear index for a full-width-at-half-maximum pulse width $t_{FWHM}$ is $n_{2,eff} = (kI_0 L)^{-1} \int_{-\infty}^{\infty} \Delta\varphi(0,t) f(0,t) dt \left( \int_{-\infty}^{\infty} f^2(0,t) dt \right)^{-1}$ where $\Delta\varphi = \Delta\varphi_\parallel(0,t) = \Delta\varphi_{elec}(0,t) + \Delta\varphi_{rot}(0,t)$ using Eq. (1), with $I(t) = I_0 f(0,t)$ and $f(0,t) = \exp(-4\ln 2\, (t/t_{FWHM})^2)$. For long pulses ($t_{FWHM} > \sim 420 fs$ for $N_2$ and $\sim 500 fs$ for $O_2$), $n_{2,eff} = n_2 + n_{2,rot}$. The values shown for $n_{2,eff}$ are maximum at the pulsewidths indicated.
(c) Pigeon et al. [16,17] measured $n_{2,eff}^{N_2} = 45 \times 10^{-20}$ cm$^2$/W and $n_{2,eff}^{O_2} = 84 \times 10^{-20}$ cm$^2$/W for a 200 ps 10.6μm pulsed $CO_2$ laser.

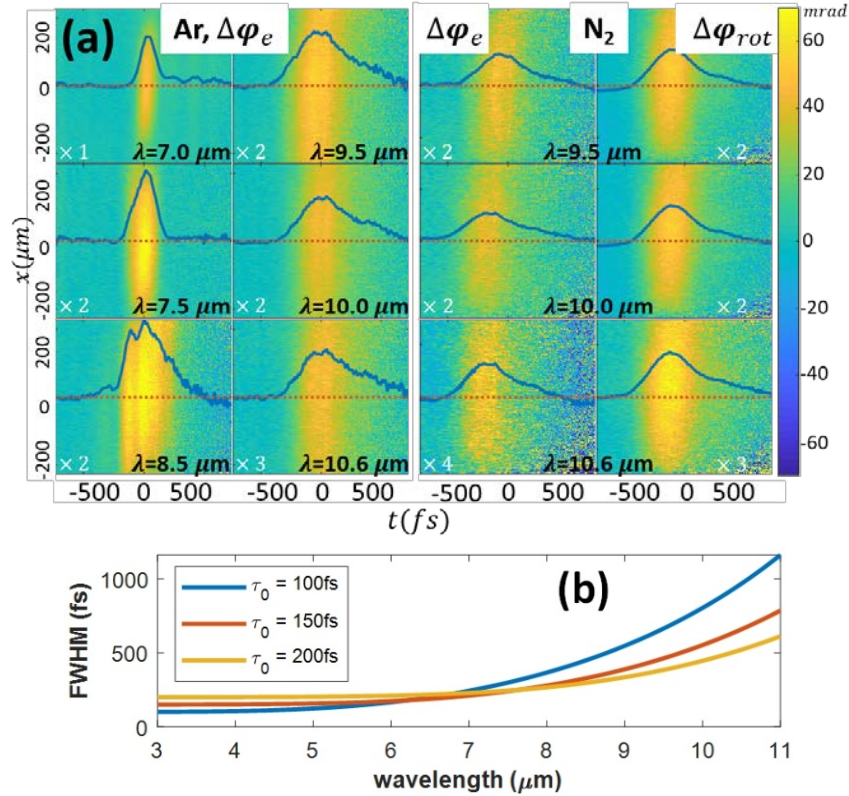

**Fig. 4.** (a) Spatiotemporal traces for λ=7.0–11µm. Left two columns: electronic nonlinear response in argon, $\Delta\varphi_{elec}(x,t) \propto I(x,t)$. Right two columns: $\Delta\varphi_{elec}(x,t)$ and $\Delta\varphi_{rot}(x,t)$ extracted from $\Delta\varphi_{\parallel}$ and $\Delta\varphi_{\perp}$. The blue curves are central lineouts. The modulations seen in $I(x,t)$ at λ=8.5µm are the result of imperfect phase matching, $\Delta \mathbf{k} = \mathbf{k}_S - \mathbf{k}_I - \mathbf{k}_{DF} \neq 0$, during that run. (b) LWIR pulse broadening vs. wavelength in $L$=9 mm of BaF$_2$ for several input pulsewidths $\tau_0$, using $\tau = \tau_0\sqrt{1 + (4ln2 \cdot GVD(\lambda_{DF}) \cdot L\tau_0^{-2})^2}$

results from the intersection volume of the signal and idler shifting in time as they propagate through the nonlinear crystal. Computing the tilt angle $\gamma$ as measured in the gas target, $\tan\gamma = -M(k_{DF}/n_{DF}) \partial\beta/\partial k_{DF}$ [23], where $\beta$ is the angle of $\mathbf{k}_{DF}$ from the crystal surface normal and $M$=45 is the pump lens demagnification factor, yields $\gamma = 7.5°$ at $\lambda = 3.0\mu m$ and $\gamma = 2.2°$ at $\lambda = 6.5\mu m$, in very good agreement with the tilt measurements in Fig. 2(a) and (b).

For small $k_{DF} = |\mathbf{k}_S - \mathbf{k}_I|$ (or long $\lambda_{DF}$), the crossing angle $\alpha$ between $\mathbf{k}_S$ and $\mathbf{k}_I$ is small, as is $k_S - k_I \approx k_{DF}$, so that the pulse front tilt is reduced at long $\lambda_{DF}$. However, temporal pulse stretching is enhanced at long $\lambda_{DF}$ owing to the increasingly negative $GVD(\lambda_{DF})$ [24] in in the BaF$_2$ focusing lens and gas cell entrance window. The calculation of Figure 4(b) shows the onset of strong broadening for $\lambda_{DF} > \sim7\mu m$, in agreement with the measurements of 4(a).

In conclusion, we have measured the electronic nonlinear index $n_2$ of the major air constituents Ar, O$_2$, and N$_2$ in the MIR-LWIR range λ=2.5-11µm and presented spatiotemporal profiles of these pulses. The values of $n_2$ show no dispersion to within our measurement precision. We observe directly, for the first time to our knowledge, the pulse front tilt and temporal broadening produced by difference frequency generation and dispersion of femtosecond infrared laser pulses.

**Funding.** Office of Naval Research (N00014-17-1-2705 and N00014-17-12778), Air Force Office of Scientific Research (FA9550-16-10284 and FA9550-16-10121).

**Acknowledgments.** The authors thank J. Wahlstrand, F. Salehi, and I. Larkin for useful discussions and technical assistance.